\documentclass[aps,two column,prl,nofootinbib,longbibliography,superscriptaddress]{revtex4-2}

\usepackage{amssymb,amsfonts}
\usepackage{amsmath}
\usepackage{mathrsfs}
\usepackage{tikz}
\usepackage{amsthm}
\usetikzlibrary{decorations.pathreplacing}

\usepackage[colorlinks, allcolors=blue!70!black, linktocpage]{hyperref}
\usepackage{accents}
\usepackage{graphicx}
\usepackage[utf8x]{inputenc}
\usepackage{float}

\usepackage{titlesec}
\usepackage{cases}
\usepackage{mathtools}
\usepackage{color}
\usepackage{cleveref}
\usepackage{braket,mathtools,float,tikz,amsfonts}
\usepackage{comment}
\usepackage{extarrows}
\newcommand{\defn}{\mathrel{\mathop:}=} %shrtct for definition operator

\definecolor{Gray}{gray}{0.95}
\definecolor{LightCyan}{rgb}{0.88,1,1}

\def\Tr{{\text{Tr}}}

\crefname{lem}{lemma}{lemmas}
\crefname{thm}{theorem}{theorems}
\crefname{cor}{corollary}{corollaries}
\crefname{rem}{remark}{remarks}
\crefname{prop}{proposition}{propositions}

\begin{document}
\count\footins = 1000

\title{Optimal bound on long-range distillable entanglement}

\author{Jonah Kudler-Flam}
\email{jkudlerflam@ias.edu}
\affiliation{School of Natural Sciences, Institute for Advanced Study, Princeton, NJ 08540, USA}
\affiliation{Princeton Center for Theoretical Science, Princeton University, Princeton, NJ 08544, USA}
\author{Vladimir Narovlansky}
\email{vladi@ias.edu}
\affiliation{School of Natural Sciences, Institute for Advanced Study, Princeton, NJ 08540, USA}
\author{Nikita Sopenko}
\email{sopenko@ias.edu}
\affiliation{School of Natural Sciences, Institute for Advanced Study, Princeton, NJ 08540, USA}

\begin{abstract}
We prove an upper bound on long-range distillable entanglement in $D$ spatial dimensions. Namely, it must decay faster than $1/r$, where $r$ is the distance between entangled regions. For states that are asymptotically rotationally invariant, the bound is strengthened to $1/r^D$. We then find explicit examples of quantum states with decay arbitrarily close to the bound. In one dimension, we construct free fermion Hamiltonians with nearest neighbor couplings that have these states as ground states. Curiously, states in conformal field theory are far from saturation, with distillable entanglement decaying faster than any polynomial.
\end{abstract}

\maketitle
% \tableofcontents

\textit{Introduction.}---Entanglement is a fundamental phenomenon that describes the intrinsically quantum correlations between subsystems. The exploration of the entanglement structure of quantum systems has had a remarkably broad impact, such as characterizing topological \cite{2006PhRvL..96k0404K,2006PhRvL..96k0405L}, gapped \cite{2007JSMTE..08...24H}, critical \cite{1994NuPhB.424..443H,calabrese2004entanglement,vidal2003entanglement}, and thermalized \cite{calabrese2005evolution} phases of matter, illuminating the emergence of spacetime \cite{ryu2006holographic,van2010building} and the black hole information problem \cite{page1993information,penington2020entanglement,almheiri2019entropy} in quantum gravity, characterizing renormalization group flows \cite{casini2004finite,casini2012renormalization}, and quantum error correction \cite{shor1995scheme} and quantum algorithms \cite{shor1994algorithms,grover1996fast}, to name just a few. 

Perhaps the most natural notion of entanglement in a (generally mixed) quantum state, $\rho_{AB}$, is the \textit{distillable entanglement}, $E_D(A,B)$ \cite{1996PhRvL..76..722B,1996PhRvA..54.3824B}. The distillable entanglement is given by the rate at which Bell pairs can be extracted from the state using only local operators and classical communications (LOCC) when $n$ copies of $\rho_{AB}$ are given while taking $n\rightarrow \infty$.
Bell pairs are the fundamental quantum resource, so this measure has an extremely clear operational interpretation. For example, once Bell pairs are distilled, they can be used for quantum teleportation \cite{bennett1993teleporting} or superdense coding \cite{bennett1992communication}. When $\rho_{AB}$ is a pure state, the distillable entanglement reduces to the entanglement entropy $E_D(A, B) = S(A) = S(B)$, though in the more generic setting of mixed states, the two quantities behave very differently.

It is natural to ask what, if any, constraints there are on the structure of distillable entanglement in many-body quantum systems. While in gapped systems, correlations decay exponentially fast, in critical systems, there can be long-range correlations as displayed by e.g.~two-point functions of local operators. We ask what sort of correlations these are, and whether genuine quantum entanglement itself can persist at long distances. Surprisingly, it was shown by Cardy, Calabrese, and Tonni \cite{calabrese2013entanglement} that the logarithmic negativity, a mixed state entanglement measure that upper bounds the distillable entanglement \cite{vidal2002computable,Plenio:2005cwa}, decays faster than polynomially in $D=1$ conformal field theories (CFT), which describe quantum critical systems. (A generalization of this result to higher $D$ \cite{Parez:2023xpj} can be understood by merging the work of \cite{calabrese2013entanglement} and \cite{cardy2013some}.) This is in contrast to the slow polynomial decay of mutual information in CFT \cite{calabrese2011entanglement,cardy2013some}, which, in addition to quantum entanglement, is also sensitive to classical correlations. Due to the fast decay of distillable entanglement in CFTs, which naively support the longest range quantum correlations, one may suspect that no quantum systems can support polynomial decay of distillable entanglement and that there is a universal upper bound. In the following, we show that there indeed is a universal upper bound on distillable entanglement. However, it is a low order polynomial, not saturated by CFTs but saturated by novel quantum states that we explicitly construct. We also provide evidence that a slow polynomial decay (which is, however, slightly faster than our bound) appears in the ground state of a spin chain with a finite-range Hamiltonian with random couplings in the phase known as a random singlet phase \cite{fisher1995critical, refael2004entanglement}. This is in sharp contrast with CFT, even though the von Neumann entropy of intervals has the same logarithmic dependence on the size.

\textit{An upper bound.}---While operationally appealing, the distillable entanglement is difficult to work with directly due to the supremum over LOCC in its definition. 
In the following, we will make use of the fact that the distillable entanglement is upper-bounded by the \textit{squashed entanglement}, which itself is defined as \cite{2004JMP....45..829C}
\begin{align}
 E_{sq}(A,B) \defn \inf_{\rho_{ABE}}\left[\frac{1}{2}I(A;B|E): \rho_{AB} = \Tr_E\rho_{ABE}\right],
\end{align}
where $I(A;B|E)\defn S(\rho_{AE}) + S(\rho_{BE})-S(\rho_{ABE}) -S(\rho_E)$ is the quantum conditional mutual information \cite{1997PhRvL..79.5194C}. The squashed entanglement is a well-behaved mixed state entanglement measure (e.g.~it is monotonically decreasing under LOCC) that crucially exhibits the ``monogamy of entanglement'' i.e.~for a tripartite Hilbert space $\mathcal{H}_A \otimes \mathcal{H}_{B_1} \otimes \mathcal{H}_{B_2}$ \cite{2004PhRvA..69b2309K}
\begin{align}
\label{eq:monogamy}
 E_{sq}(A,B_1) + E_{sq}(A, B_2) \leq E_{sq}(A, B_1 \cup B_2).
\end{align}
That is, if two systems are maximally entangled, they are unable to be entangled with any other system. This has no classical analog, as classical systems are polygamous; they can be statistically correlated with many different other systems at the same time. Consequently, measures of classical correlation, such as mutual information, need not decay at large distances. This can be easily seen in the GHZ state \cite{greenberger1989going}
\begin{align}
 \ket{GHZ} = \frac{1}{\sqrt{2}}\left(\ket{000\dots} + \ket{111\dots}\right),
\end{align}
where the mutual information between any two qubits is $\log 2$ while the distillable entanglement is always $0$. The zero-mode contribution to the vacuum state of a free massless scalar was argued to have GHZ-like behavior in \cite{2011arXiv1107.2940H}, giving some insight on the difference between classical and quantum correlations in CFT.

\begin{figure}
 \centering
 \begin{tikzpicture}[scale=1]
\foreach \i in {1,...,120}
{
\filldraw[black] (\i/16,0) circle (.25pt);
}
\draw[line width=1.5mm, blue, opacity = .7] (.9,0) -- (2.1,0);
\draw[line width=1.5mm,, red, opacity = .7] (3.9,0) -- (4.6,0);
\draw[line width=1.5mm,, red, opacity = .7] (4.65,0) -- (5.35,0);
\draw[line width=1.5mm,, red, opacity = .7] (5.4,0) -- (6.1,0);
\draw[line width=1.5mm,, red, opacity = .7] (6.15,0) -- (6.85,0);
\node[] at (1.5,-.35) {$A$};
\node[] at (4.25,-.35) {$B_1$};
\node[] at (5,-.35) {$B_2$};
\node[] at (5.75,-.35) {$B_3$};
\node[] at (6.5,-.35) {$B_4$};
\node[] at (7.1,-.35) {$\dots$};
\end{tikzpicture}
 \caption{A one-dimensional chain of qudits with subsystems $B_i$ (red) successively becoming further away from subsystem $A$ (green).}
 \label{fig:chain}
\end{figure}
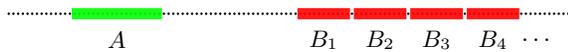

Consider a one-dimensional chain of qudits, each with Hilbert space dimension $d$ (see Figure \ref{fig:chain}). Taking $\mathcal{H}_A$ to be a subsystem of $N_A$ contiguous qudits, the squashed entanglement of $A$ with any other subsystem, $C$, is bounded as
\begin{align}
\label{maxent_bound}
 E_{sq}(A,C) \leq N_A \log d < \infty. 
\end{align}
Taking contiguous subregions $B_i$, each with $N_{B_i}$ qudits, the monogamy inequality may be iterated to the multipartite statement
\begin{align}
 \sum_i E_{sq}(A,B_i) \leq E_{sq}(A, \cup_i B_i).
\end{align}
We may view $E_{sq}(A,B_i)$ as a function $f(r)$ of the distance, $r$, between $A$ and $B_i$. Assuming that we do not take $N_{B_i}$ to grow with $r$, we get the convergence condition $\sum_{r=0}^{\infty} f(r) < \infty$ on the sequence $\{f(r)\}_{r \in \mathbb{N}}$. In particular, it excludes the possibility of slow polynomial decay $\sim r^{-\alpha}$ of $f(r)$ for $\alpha \leq 1$\footnote{More precisely, the sequence $\{f(r)\}_{r=1}^{\infty}$ cannot be lower bounded by $C r^{-\alpha}$ for any $C>0$ and $\alpha \leq 1$.}. We note that in this argument, we have not used any special properties that the quantum state may have, e.g.~locality or translational invariance.
% Assuming that we do not take $N_{B_i}$ to grow with $r$, the sum only converges if $E_{sq}(A,B_i)$, and consequently the distillable entanglement, decays strictly faster than $1/r$ (on average) as $r\rightarrow \infty$. By ``on average,'' we mean that the partial sums are bounded by the partial sums of $C/r$ for some constant $C$. Technically, convergence requires a condition slightly stronger than this because one can include logarithmic corrections to the power law. We will colloquially refer to this as $1/r$ behavior, with the understanding that these corrections can be easily incorporated. We note that in this argument, we have not used any special properties that the quantum state may have, e.g.~locality or translational invariance.

In higher spatial dimensions, $D$, there is a subtlety when deriving a bound. Considering a one-dimensional subsystem of the system and the distillable entanglement as a function of $r$ in this subsystem, the previous argument implies the decay must be faster than $1/r$. As we will soon show, we can construct one-dimensional states with $1/r^{1+\epsilon}$ decay for any $\epsilon > 0$, so simply taking the tensor product of this state and any product state on the $D$-dimensional system gets arbitrarily close to the $1/r$ bound in $D$ dimensions. However, this state is quite fine-tuned because it has no entanglement in the transverse directions, so we are able to make a stronger statement with physical assumptions.

\begin{figure}
 \centering
\begin{tikzpicture}[scale=1]
\foreach \i in {2,...,99}
{
\foreach \j in {2,...,99}
{
\filldraw[gray] (\i/16,\j/16) circle (.25pt);
}
}
\filldraw[blue,opacity = .7] (3.15,3.15) circle (15pt);
\fill [red,even odd rule,opacity = .7] (3.15,3.15) circle[radius=2cm] circle[radius=1.6cm];
\fill [red,even odd rule,opacity = .7] (3.15,3.15) circle[radius=2.5cm] circle[radius=2.1cm];
\fill [red,even odd rule,opacity = .7] (3.15,3.15) circle[radius=3cm] circle[radius=2.6cm];
\node[] at (3.15,3.15){$A$};
\node[] at (4.95,3.15){$B_1$};
\node[] at (5.45,3.15){$B_2$};
\node[] at (5.95,3.15){$B_3$};
\draw[very thick, <->] (0.15,3.15) -- (.525,3.15);
\node[] at (0.35,3.4) {$\Delta r$}; 
\end{tikzpicture}
 \caption{A two-dimensional lattice of qudits with concentric annuli of width $\Delta r$ becoming further away from subsystem $A$.}
 \label{fig:concentric}
\end{figure}
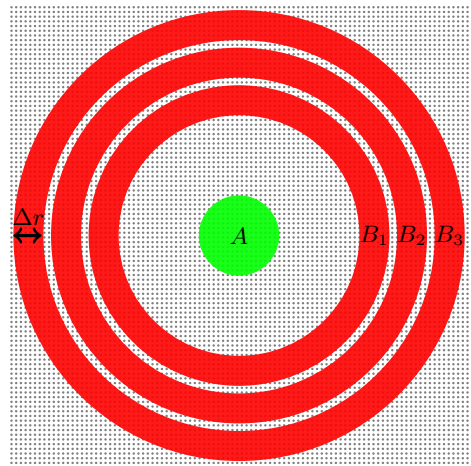

Consider $B_i$ to be concentric subregions of characteristic width $\Delta r$ (see Figure \ref{fig:concentric}). The number of qudits in the subregion grows as $r^{D-1}$. Applying the monogamy of $E_{sq}(A,B_i)$ inequality leads to a decay faster than $1/r$ even though the Hilbert space dimension of $B_i$ is now growing with $r$ (something we did not allow for $D=1$). Breaking each $B_i$ into subregions that have a fixed number of qudits, the average distillable entanglement of these subregions must then decay faster than $1/r^D$. If the quantum state is asymptotically rotationally invariant at large $r$, the distillable entanglement must decay faster than $1/r^D$ for any disk-like subregions. 

\textit{Optimality.}---Given that conformal field theories have long range distillable entanglement that decays faster than any polynomial, the upper bounds that we derived may appear to be very weak. We now describe explicit classes of states of spin chains that have long range decay of entanglement as $1/r^{1+\epsilon}$ for any $\epsilon > 0$, showing the optimality of our bound. We also argue for the existence of asymptotically rotationally invariant states of $D$-dimensional spin systems for which we get a $1/r^{D+\epsilon}$ decay.

In order to approach the bound on long distance entanglement, we are inspired to consider a simple class of states that saturate the monogamy of entanglement inequality. It is easy to see that any state that is a tensor product of maximally entangled ($d$-dimensional) Bell pairs and locally maximally mixed states saturates \eqref{eq:monogamy}. Both the squashed and distillable entanglements for two subregions are simply given by the number of Bell pairs connecting the regions (times $\log d$).

\begin{figure}
    \centering
    \begin{tikzpicture}[scale=1]
\foreach \i in {1,...,40}
{
\filldraw[black] (\i/5,0) circle (.5pt);
}
\draw[black] (.8,0) arc[start angle=0, end angle=180, radius=0.3];
\draw[black] (.6,0) arc[start angle=0, end angle=180, radius=0.1];

\draw[black] (2.4,0) arc[start angle=0, end angle=180, radius=0.7];
\draw[black] (2.2,0) arc[start angle=0, end angle=180, radius=0.5];
\draw[black] (2,0) arc[start angle=0, end angle=180, radius=0.3];
\draw[black] (1.8,0) arc[start angle=0, end angle=180, radius=0.1];

\draw[black] (2.8,0) arc[start angle=0, end angle=180, radius=0.1];

\draw[black] (6.4,0) arc[start angle=0, end angle=180, radius=1.7];
\draw[black] (6.2,0) arc[start angle=0, end angle=180, radius=1.5];
\draw[black] (6,0) arc[start angle=0, end angle=180, radius=1.3];
\draw[black] (5.8,0) arc[start angle=0, end angle=180, radius=1.1];
\draw[black] (5.6,0) arc[start angle=0, end angle=180, radius=0.9];
\draw[black] (5.4,0) arc[start angle=0, end angle=180, radius=0.7];
\draw[black] (5.2,0) arc[start angle=0, end angle=180, radius=0.5];
\draw[black] (5,0) arc[start angle=0, end angle=180, radius=0.3];
\draw[black] (4.8,0) arc[start angle=0, end angle=180, radius=0.1];

\draw[black] (8,0) arc[start angle=0, end angle=180, radius=.5];
\draw[black] (7.8,0) arc[start angle=0, end angle=180, radius=.3];
\draw[black] (7.6,0) arc[start angle=0, end angle=180, radius=.1];

\draw[black] (6.8,0) arc[start angle=0, end angle=180, radius=.1];

\node[] at (.5,-.4) {$2k_1$};
\node[] at (1.7,-.4) {$2k_2$};
\node[] at (2.7,-.4) {$2k_3$};
\node[] at (4.7,-.4) {$2k_4$};
\node[] at (6.7,-.4) {$2k_5$};
\node[] at (7.5,-.4) {$2k_6$};

\draw[decorate, decoration={brace, amplitude=3pt, mirror}] (0.2,-.1) -- (.8,-.1);
\draw[decorate, decoration={brace, amplitude=3pt, mirror}] (1,-.1) -- (2.4,-.1);
\draw[decorate, decoration={brace, amplitude=3pt, mirror}] (2.6,-.1) -- (2.8,-.1);
\draw[decorate, decoration={brace, amplitude=3pt, mirror}] (3,-.1) -- (6.4,-.1);
\draw[decorate, decoration={brace, amplitude=3pt, mirror}] (6.6,-.1) -- (6.8,-.1);
\draw[decorate, decoration={brace, amplitude=3pt, mirror}] (7,-.1) -- (8,-.1);
\end{tikzpicture}
    \caption{An example of a portion of the chain with fragments composed of rainbows of sizes drawn from a distribution.}
    \label{fig:rainbows_tikz}
\end{figure}
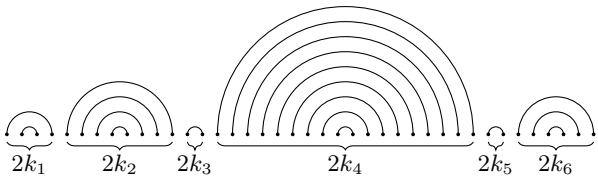

We consider a (normalized) probability distribution of the distance connecting Bell pairs (in one spatial dimension) given at large $r$ as
\begin{align}
 P_{\alpha}(r) = C_{\alpha,1}{r^{-\alpha}}.
\end{align}
Here we measure distances in units of the size of a qudit. This is a well-defined probability distribution when $\alpha > 1$, and one can find a pure state of an infinite spin chain realizing it in the following way (see Figure \ref{fig:rainbows_tikz}). We first generate a sequence of positive integers $\{k_i\}_{i \in \mathbb{Z}}$ with probability distribution $\sim 1/k^{1+\alpha}$. We then divide the infinite spin chain of qubits into fragments of length $2 k_i$ and consider a state in which $n$-th qubit of the $i$-th fragment is maximally entangled with $(2k_i+1-n)$-th qubit (of the same fragment). In the resulting state, the number of Bell pairs of length $(2n+1)$ in a large interval of length $L$ is given by \begin{align}
    \sim \frac{L}{\sum_{k=1}^{\infty} k^{-1-\alpha}} \sum_{k > n}^{\infty} 1/k^{1+\alpha} \sim L/(n+1)^{\alpha}
\end{align}
which gives the scaling $L/r^{\alpha}$ as $r \to \infty$.

 % For a finite number of qudits, one can consider a distribution that approximates a polynomial decay with $\alpha \leq 1$. However, when taking the thermodynamic limit, this will lead to a state where the probability of any two finite subregions being entangled to be zero, trivially satisfying our bound.

 We note two interesting scenarios, previously studied in the literature, where this probability distribution appears: (1) The random singlet phase \cite{refael2004entanglement}, which is the fixed point of the strong disorder renormalization group \cite{fisher1995critical} has a ground state that qualitatively can be described by a collection of Bell pairs with polynomially decaying scaling for $\alpha = 2$ at large $r$. We will return to this example shortly and discuss a possible generalization for any $\alpha > 1$. (2) At the critical point of monitored Clifford quantum circuits, it has been numerically observed that $\alpha \simeq 6$ at large $r$ \cite{shi2020entanglement}.\footnote{We also note that polynomial decay of entanglement has been noted to be common in fermionic systems \cite{2024arXiv240206677P}.}

Let us first consider the entanglement entropy for a region of $N_A$ qudits. The entanglement entropy is given by the number of Bell pairs that connect the region to its complement (times $\log d$). This is
\begin{align}
 \overline{S}(\alpha ,N_A) = \sum_{i = 1}^{N_A} \left[\frac{1}{2} \sum_{r=i}^{\infty}P_{\alpha}(r) + \frac{1}{2} \sum_{r= N_A-i+1}^{\infty}P_{\alpha}(r) \right] \log d,
\end{align}
where the overline denotes ensemble averaging. The two terms represent the equal probabilities that the $i^{th}$ qudit in $N_A$ is entangled with a qudit located to its left or right.
When $N_A$ is large, we may approximate the sums by integrals. Furthermore, this is the regime where the ensemble average is a good approximation to individual realizations
\begin{align}
\label{eq:ents}
\begin{aligned}
 S(\alpha ,N_A) 
 % &\simeq \frac{\zeta (\alpha -1,l)-\zeta (\alpha -1)}{\zeta (\alpha )-\alpha \zeta (\alpha )}\log d
 % \\
 &\simeq \begin{cases}
 \frac{C_{\alpha,1}\log d}{3 \alpha -\alpha ^2-2}N_A^{2-\alpha }, & 1 < \alpha < 2
 \\
 \\
 {C_{\alpha,1}\log d} \log N_A,& \alpha = 2
 \\
 \\
 \frac{C_{\alpha,1}\log d}{\alpha ^2-3 \alpha +2}, & \alpha > 2
 \end{cases}
\end{aligned}.
\end{align}
We therefore find fractional power law growth for $1 < \alpha <2$, area law entanglement for $\alpha > 2$ and logarithmic growth at the critical point $\alpha = 2$. The area law coefficient will generically depend on the small $r$ behavior of the probability distribution.

For the distillable entanglement of disjoint intervals, with $N_A$ and $N_B$ qudits respectively, and separated by a distance $r$, 
\begin{align}
 \overline{E_D}(\alpha, N_A,N_B,r) &= \sum_{i = 1}^{N_A}\frac{\log d}{2}\sum_{r' = r+i}^{r+i+N_B-1}P_{\alpha}(r') .
\end{align}
When $r \gg N_A,N_B \gg 1$
\begin{align}
 \overline{E_D}(\alpha, N_A,N_B,r)\simeq \frac{ C_{\alpha,1}\log d}{2 } N_A N_Br^{-\alpha }.
\end{align}
The distillable entanglement has large fluctuations in this regime, so we have not removed the overline.
Therefore, we are able to achieve a long distance polynomial decay for any $\alpha > 1$. The convergence of the sum of $E_{sq}(A,B_i)$'s in our derivation in the bound is directly connected to the integrability of the probability distribution for distances between Bell pairs.

In $D$ spatial dimensions, we consider a rotationally symmetric probability distribution at large $r$ to be
\begin{align}
 P_{\alpha}(r) =C_{\alpha,D} {r^{-\alpha}},
\end{align}
where $C_{\alpha,D}$ is a normalization constant.
This probability distribution is normalizable when $\alpha > 1$. This distribution leads to a (on average) rotationally invariant state. We consider two subsystems of $N_A$ and $N_B$ qudits and separated by a distance $r$. For $1 \ll N_A,N_B \ll r$, the entanglement is given by
\begin{align}
 \overline{E_D}(\alpha,N_A,N_B,r) \simeq \frac{C_{\alpha,D}N_AN_B \log d}{\text{Area}_{S_{D-1}}} ~r^{1-D-\alpha},
\end{align}
where $\text{Area}_{S_{D-1}}$ is the area of the unit $(D-1)$-sphere. 
Here we work in units where the qudit density is 1.
This saturates our bound when $\alpha$ approaches $1$.

\textit{Free fermions.}---
We now explicitly study a local quantum system exhibiting polynomial decay of distillable entanglement. A state given by a collection of Bell pairs on a one-dimensional lattice with distances between pairs distributed by $P(r) = C r^{-2}$ (for large $r$) provides a qualitative picture of the ground state of the random singlet phase \cite{refael2004entanglement}. It suggests that the distillable entanglement of two intervals $A$, $B$ decays polynomially as $1/r^2$ with the distance between intervals. We give numerical evidence that it is actually the case by studying the ground state of a free fermionic system
\begin{align}
H = \frac{1}{2} \sum_{j \in \mathbb{Z}} J_{j,j+1} \left( c^{\dagger}_j c_{j+1} + c^{\dagger}_{j+1} c_{j} \right)
\end{align}
with $J_{j,j+1}$ chosen randomly from $[0,1]$ with the probability distribution $P(J) = \frac{1}{\delta} J^{-1+1/\delta}$. This probability distribution appears naturally in the strong disorder renormalization group analysis with the parameter $\delta$ measuring the strength of the disorder \cite{fisher1995critical}. The logarithmic growth of entanglement entropy, as seen in \eqref{eq:ents}, has been numerically observed in \cite{2005PhRvB..72n0408L}.

Recall that for two subsystems $A$, $B$ the coherent mutual information from $A$ to $B$ is defined by $I_c(A \rangle B) := S(B) - S(A \cup B)$. The hashing inequality states that it lower bounds the distillable entanglement entropy $I_c(A \rangle B) \leq E_D(A,B)$ \cite{1996PhRvA..54.3824B,2005RSPSA.461..207D,2000PhRvL..85..433H}. We let $I_c(A,B)$ be the maximum between $I_c(A \rangle B)$, $I_c(B \rangle A)$ and 0.

 \begin{figure}
 \centering
 \includegraphics[width=0.45\textwidth]{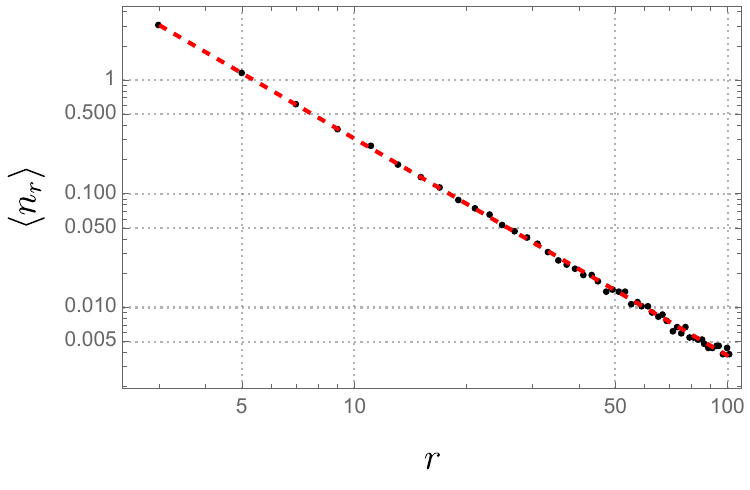}
 \caption{The dependence of the average number of approximate Bell pairs with the coherent mutual information greater than $\log(2)/2$ on the distance between the sites in a pair. The data, which may be found here \cite{Long_Range_Entanglement_Figure4_2025}, is given for $N=30500$ samples of a free fermionic spin chain of size $L=500$ with the distribution $P(J)$ for $\delta = 3$. The polynomial fit for $r \in [3,101]$ gives the exponent $\alpha = 1.912$. The fit in the intermediate range of $r$ gives $\alpha$ inside the interval $[1.8, 2.2]$.}
 \label{fig:RSPgraph}
\end{figure}
Let $A$ and $B$ be two intervals with decompositions $A = \bigcup_{i \in S} A_i$, $B = \bigcup_{i \in S} B_i$ into disjoint subsets labeled by a set $S$. Using the superadditivity of the distillable entanglement and the hashing inequality, we have 
\begin{align}
 E_D(A,B) \geq \sum_{i \in S} E_D(A_i,B_i) \geq \sum_{i \in S} I_c(A_i,B_i). 
\end{align}
Thus, in order to show that $E_D(A,B)$ decays at most as $\sim 1/r^2$ it is enough to check that we have pairs of sites $j \in A$, $k \in B$ with sufficiently large (greater than some fixed constant) coherent mutual information. For that, we study the distribution of pairs of sites in a randomly generated ground state which are sufficiently entangled with each other having\footnote{The threshold value $\tfrac12 \log 2$ has been chosen arbitrarily. We find a similar behavior for other values.} $I_c(\{j\},\{k\}) > \tfrac{1}{2} \log 2$. We find that the fraction of such pairs does not depend on the system size, and their distribution is well approximated by $P(r) = C r^{-\alpha}$ for $\alpha$ close to 2 (see Fig. \ref{fig:RSPgraph}). 

We now describe a generalization to accommodate any $\alpha > 1$. To do so, we first describe how to create a ``rainbow'' of $k_i$ singlets in a fragment of $2N=2k_i$ spins via a local Hamiltonian. To do so, we take $J_{N,N+1} = 1$ and $J_{j,j+1} = \alpha^{2|N-j|-1}$ for $j=1, ..., 2N$ and $j \neq N$ and $0<\alpha<1$. This is the coupling of the so-called ``rainbow chain'' \cite{2010NJPh...12k3049V,2014JSMTE..10..004R, 2015JSMTE..06..002R,2017JPhA...50p4001R}. We check numerically that the coherent mutual information between qubits $i$ and $(2N+1-i)$ for $i=1,..., N$ is bounded from below by some positive constant that depends on $\alpha$ but independent of the system size. We then string together a sequence of fragments with probability distribution $\sim 1/k^{1+\alpha}$, which we previously argued gives the desired distribution of Bell pairs. The coupling between rainbows is taken to either be zero or parametrically small so that the global ground state is the tensor product of the ground states of the fragments. While only nearest-neighbor interactions are included in this Hamiltonian, the correlations in the coupling are non-local, making this model somewhat less natural than the random singlet phase.

\textit{Discussion.}---We have shown that there is a universal upper bound on the long-distance distillable entanglement present in quantum states in any dimension. Depending on the assumption of asymptotic rotational symmetry, this bound is dimension dependent. We then demonstrated via explicit construction of quantum states that the bound is optimal. The crucial ingredient that allowed for such a bound was the monogamy of quantum entanglement. Without such a restriction, no bound can be made, as we demonstrated for classical correlations via the GHZ state. The monogamy of entanglement simply led to an integrability condition on the distribution of distillable entanglement.

At no point in our derivation of the optimal bound did we refer to any Hamiltonian or physical conditions set on that Hamiltonian, such as locality of interactions or translational invariance. It would be particularly interesting if natural states, such as the ground state, of physical Hamiltonians can saturate or approach the bound. Of course, this will not be the case for theories that are conformally invariant because the decay is faster than polynomial. Moreover, Lorentz invariance implies that $\partial_{N_A} (N_A S(N_A)) \leq 0$ \cite{casini2004finite}, so the entanglement scaling in \eqref{eq:ents} for $1 < \alpha < 2$ (which in our examples led to the saturation of the distillable entanglement) is disallowed for Lorentz invariant states. Perhaps deformations of the random singlet phase, monitored quantum circuits, or the ``Motzkin chain,'' which has anomalously large entanglement \cite{movassagh2016supercritical,bravyi2012criticality,zhang2016quantum}, can provide clues.

\begin{acknowledgements}
 \textit{Acknowledgments.}---We thank Matt Headrick, Aidan Herderschee, and Shinsei Ryu for discussions. We especially thank Juan Maldacena for discussions that prompted this work.
 JKF is supported by the Marvin L. Goldberger Member Fund at the Institute for Advanced Study.
NS is supported by the Ambrose Monell
Foundation.
 JKF, VN, and NS gratefully acknowledge support from NSF Grant PHY-2207584.
\end{acknowledgements}

\bibliographystyle{apsrev}
\bibliography{main}

\begin{thebibliography}{47}
\expandafter\ifx\csname natexlab\endcsname\relax\def\natexlab#1{#1}\fi
\expandafter\ifx\csname bibnamefont\endcsname\relax
  \def\bibnamefont#1{#1}\fi
\expandafter\ifx\csname bibfnamefont\endcsname\relax
  \def\bibfnamefont#1{#1}\fi
\expandafter\ifx\csname citenamefont\endcsname\relax
  \def\citenamefont#1{#1}\fi
\expandafter\ifx\csname url\endcsname\relax
  \def\url#1{\texttt{#1}}\fi
\expandafter\ifx\csname urlprefix\endcsname\relax\def\urlprefix{URL }\fi
\providecommand{\bibinfo}[2]{#2}
\providecommand{\eprint}[2][]{\url{#2}}

\bibitem[{\citenamefont{{Kitaev} and {Preskill}}(2006)}]{2006PhRvL..96k0404K}
\bibinfo{author}{\bibfnamefont{A.}~\bibnamefont{{Kitaev}}} \bibnamefont{and} \bibinfo{author}{\bibfnamefont{J.}~\bibnamefont{{Preskill}}}, \bibinfo{journal}{\prl} \textbf{\bibinfo{volume}{96}}, \bibinfo{eid}{110404} (\bibinfo{year}{2006}), \eprint{hep-th/0510092}.

\bibitem[{\citenamefont{{Levin} and {Wen}}(2006)}]{2006PhRvL..96k0405L}
\bibinfo{author}{\bibfnamefont{M.}~\bibnamefont{{Levin}}} \bibnamefont{and} \bibinfo{author}{\bibfnamefont{X.-G.} \bibnamefont{{Wen}}}, \bibinfo{journal}{\prl} \textbf{\bibinfo{volume}{96}}, \bibinfo{eid}{110405} (\bibinfo{year}{2006}), \eprint{cond-mat/0510613}.

\bibitem[{\citenamefont{{Hastings}}(2007)}]{2007JSMTE..08...24H}
\bibinfo{author}{\bibfnamefont{M.~B.} \bibnamefont{{Hastings}}}, \bibinfo{journal}{Journal of Statistical Mechanics: Theory and Experiment} \textbf{\bibinfo{volume}{2007}}, \bibinfo{pages}{08024} (\bibinfo{year}{2007}), \eprint{0705.2024}.

\bibitem[{\citenamefont{{Holzhey} et~al.}(1994)\citenamefont{{Holzhey}, {Larsen}, and {Wilczek}}}]{1994NuPhB.424..443H}
\bibinfo{author}{\bibfnamefont{C.}~\bibnamefont{{Holzhey}}}, \bibinfo{author}{\bibfnamefont{F.}~\bibnamefont{{Larsen}}}, \bibnamefont{and} \bibinfo{author}{\bibfnamefont{F.}~\bibnamefont{{Wilczek}}}, \bibinfo{journal}{Nuclear Physics B} \textbf{\bibinfo{volume}{424}}, \bibinfo{pages}{443} (\bibinfo{year}{1994}), \eprint{hep-th/9403108}.

\bibitem[{\citenamefont{Calabrese and Cardy}(2004)}]{calabrese2004entanglement}
\bibinfo{author}{\bibfnamefont{P.}~\bibnamefont{Calabrese}} \bibnamefont{and} \bibinfo{author}{\bibfnamefont{J.}~\bibnamefont{Cardy}}, \bibinfo{journal}{Journal of statistical mechanics: theory and experiment} \textbf{\bibinfo{volume}{2004}}, \bibinfo{pages}{P06002} (\bibinfo{year}{2004}).

\bibitem[{\citenamefont{Vidal et~al.}(2003)\citenamefont{Vidal, Latorre, Rico, and Kitaev}}]{vidal2003entanglement}
\bibinfo{author}{\bibfnamefont{G.}~\bibnamefont{Vidal}}, \bibinfo{author}{\bibfnamefont{J.~I.} \bibnamefont{Latorre}}, \bibinfo{author}{\bibfnamefont{E.}~\bibnamefont{Rico}}, \bibnamefont{and} \bibinfo{author}{\bibfnamefont{A.}~\bibnamefont{Kitaev}}, \bibinfo{journal}{Physical review letters} \textbf{\bibinfo{volume}{90}}, \bibinfo{pages}{227902} (\bibinfo{year}{2003}).

\bibitem[{\citenamefont{Calabrese and Cardy}(2005)}]{calabrese2005evolution}
\bibinfo{author}{\bibfnamefont{P.}~\bibnamefont{Calabrese}} \bibnamefont{and} \bibinfo{author}{\bibfnamefont{J.}~\bibnamefont{Cardy}}, \bibinfo{journal}{Journal of Statistical Mechanics: Theory and Experiment} \textbf{\bibinfo{volume}{2005}}, \bibinfo{pages}{P04010} (\bibinfo{year}{2005}).

\bibitem[{\citenamefont{Ryu and Takayanagi}(2006)}]{ryu2006holographic}
\bibinfo{author}{\bibfnamefont{S.}~\bibnamefont{Ryu}} \bibnamefont{and} \bibinfo{author}{\bibfnamefont{T.}~\bibnamefont{Takayanagi}}, \bibinfo{journal}{Physical review letters} \textbf{\bibinfo{volume}{96}}, \bibinfo{pages}{181602} (\bibinfo{year}{2006}).

\bibitem[{\citenamefont{Van~Raamsdonk}(2010)}]{van2010building}
\bibinfo{author}{\bibfnamefont{M.}~\bibnamefont{Van~Raamsdonk}}, \bibinfo{journal}{International Journal of Modern Physics D} \textbf{\bibinfo{volume}{19}}, \bibinfo{pages}{2429} (\bibinfo{year}{2010}).

\bibitem[{\citenamefont{Page}(1993)}]{page1993information}
\bibinfo{author}{\bibfnamefont{D.~N.} \bibnamefont{Page}}, \bibinfo{journal}{Physical review letters} \textbf{\bibinfo{volume}{71}}, \bibinfo{pages}{3743} (\bibinfo{year}{1993}).

\bibitem[{\citenamefont{Penington}(2020)}]{penington2020entanglement}
\bibinfo{author}{\bibfnamefont{G.}~\bibnamefont{Penington}}, \bibinfo{journal}{Journal of High Energy Physics} \textbf{\bibinfo{volume}{2020}}, \bibinfo{pages}{1} (\bibinfo{year}{2020}).

\bibitem[{\citenamefont{Almheiri et~al.}(2019)\citenamefont{Almheiri, Engelhardt, Marolf, and Maxfield}}]{almheiri2019entropy}
\bibinfo{author}{\bibfnamefont{A.}~\bibnamefont{Almheiri}}, \bibinfo{author}{\bibfnamefont{N.}~\bibnamefont{Engelhardt}}, \bibinfo{author}{\bibfnamefont{D.}~\bibnamefont{Marolf}}, \bibnamefont{and} \bibinfo{author}{\bibfnamefont{H.}~\bibnamefont{Maxfield}}, \bibinfo{journal}{Journal of High Energy Physics} \textbf{\bibinfo{volume}{2019}}, \bibinfo{pages}{1} (\bibinfo{year}{2019}).

\bibitem[{\citenamefont{Casini and Huerta}(2004)}]{casini2004finite}
\bibinfo{author}{\bibfnamefont{H.}~\bibnamefont{Casini}} \bibnamefont{and} \bibinfo{author}{\bibfnamefont{M.}~\bibnamefont{Huerta}}, \bibinfo{journal}{Physics Letters B} \textbf{\bibinfo{volume}{600}}, \bibinfo{pages}{142} (\bibinfo{year}{2004}).

\bibitem[{\citenamefont{Casini and Huerta}(2012)}]{casini2012renormalization}
\bibinfo{author}{\bibfnamefont{H.}~\bibnamefont{Casini}} \bibnamefont{and} \bibinfo{author}{\bibfnamefont{M.}~\bibnamefont{Huerta}}, \bibinfo{journal}{Physical Review D—Particles, Fields, Gravitation, and Cosmology} \textbf{\bibinfo{volume}{85}}, \bibinfo{pages}{125016} (\bibinfo{year}{2012}).

\bibitem[{\citenamefont{Shor}(1995)}]{shor1995scheme}
\bibinfo{author}{\bibfnamefont{P.~W.} \bibnamefont{Shor}}, \bibinfo{journal}{Physical review A} \textbf{\bibinfo{volume}{52}}, \bibinfo{pages}{R2493} (\bibinfo{year}{1995}).

\bibitem[{\citenamefont{Shor}(1994)}]{shor1994algorithms}
\bibinfo{author}{\bibfnamefont{P.~W.} \bibnamefont{Shor}}, in \emph{\bibinfo{booktitle}{Proceedings 35th annual symposium on foundations of computer science}} (\bibinfo{organization}{Ieee}, \bibinfo{year}{1994}), pp. \bibinfo{pages}{124--134}.

\bibitem[{\citenamefont{Grover}(1996)}]{grover1996fast}
\bibinfo{author}{\bibfnamefont{L.~K.} \bibnamefont{Grover}}, in \emph{\bibinfo{booktitle}{Proceedings of the twenty-eighth annual ACM symposium on Theory of computing}} (\bibinfo{year}{1996}), pp. \bibinfo{pages}{212--219}.

\bibitem[{\citenamefont{{Bennett} et~al.}(1996{\natexlab{a}})\citenamefont{{Bennett}, {Brassard}, {Popescu}, {Schumacher}, {Smolin}, and {Wootters}}}]{1996PhRvL..76..722B}
\bibinfo{author}{\bibfnamefont{C.~H.} \bibnamefont{{Bennett}}}, \bibinfo{author}{\bibfnamefont{G.}~\bibnamefont{{Brassard}}}, \bibinfo{author}{\bibfnamefont{S.}~\bibnamefont{{Popescu}}}, \bibinfo{author}{\bibfnamefont{B.}~\bibnamefont{{Schumacher}}}, \bibinfo{author}{\bibfnamefont{J.~A.} \bibnamefont{{Smolin}}}, \bibnamefont{and} \bibinfo{author}{\bibfnamefont{W.~K.} \bibnamefont{{Wootters}}}, \bibinfo{journal}{\prl} \textbf{\bibinfo{volume}{76}}, \bibinfo{pages}{722} (\bibinfo{year}{1996}{\natexlab{a}}), \eprint{quant-ph/9511027}.

\bibitem[{\citenamefont{{Bennett} et~al.}(1996{\natexlab{b}})\citenamefont{{Bennett}, {Divincenzo}, {Smolin}, and {Wootters}}}]{1996PhRvA..54.3824B}
\bibinfo{author}{\bibfnamefont{C.~H.} \bibnamefont{{Bennett}}}, \bibinfo{author}{\bibfnamefont{D.~P.} \bibnamefont{{Divincenzo}}}, \bibinfo{author}{\bibfnamefont{J.~A.} \bibnamefont{{Smolin}}}, \bibnamefont{and} \bibinfo{author}{\bibfnamefont{W.~K.} \bibnamefont{{Wootters}}}, \bibinfo{journal}{\pra} \textbf{\bibinfo{volume}{54}}, \bibinfo{pages}{3824} (\bibinfo{year}{1996}{\natexlab{b}}), \eprint{quant-ph/9604024}.

\bibitem[{\citenamefont{Bennett et~al.}(1993)\citenamefont{Bennett, Brassard, Cr{\'e}peau, Jozsa, Peres, and Wootters}}]{bennett1993teleporting}
\bibinfo{author}{\bibfnamefont{C.~H.} \bibnamefont{Bennett}}, \bibinfo{author}{\bibfnamefont{G.}~\bibnamefont{Brassard}}, \bibinfo{author}{\bibfnamefont{C.}~\bibnamefont{Cr{\'e}peau}}, \bibinfo{author}{\bibfnamefont{R.}~\bibnamefont{Jozsa}}, \bibinfo{author}{\bibfnamefont{A.}~\bibnamefont{Peres}}, \bibnamefont{and} \bibinfo{author}{\bibfnamefont{W.~K.} \bibnamefont{Wootters}}, \bibinfo{journal}{Physical review letters} \textbf{\bibinfo{volume}{70}}, \bibinfo{pages}{1895} (\bibinfo{year}{1993}).

\bibitem[{\citenamefont{Bennett and Wiesner}(1992)}]{bennett1992communication}
\bibinfo{author}{\bibfnamefont{C.~H.} \bibnamefont{Bennett}} \bibnamefont{and} \bibinfo{author}{\bibfnamefont{S.~J.} \bibnamefont{Wiesner}}, \bibinfo{journal}{Physical review letters} \textbf{\bibinfo{volume}{69}}, \bibinfo{pages}{2881} (\bibinfo{year}{1992}).

\bibitem[{\citenamefont{Calabrese et~al.}(2013)\citenamefont{Calabrese, Cardy, and Tonni}}]{calabrese2013entanglement}
\bibinfo{author}{\bibfnamefont{P.}~\bibnamefont{Calabrese}}, \bibinfo{author}{\bibfnamefont{J.}~\bibnamefont{Cardy}}, \bibnamefont{and} \bibinfo{author}{\bibfnamefont{E.}~\bibnamefont{Tonni}}, \bibinfo{journal}{Journal of Statistical Mechanics: Theory and Experiment} \textbf{\bibinfo{volume}{2013}}, \bibinfo{pages}{P02008} (\bibinfo{year}{2013}).

\bibitem[{\citenamefont{Vidal and Werner}(2002)}]{vidal2002computable}
\bibinfo{author}{\bibfnamefont{G.}~\bibnamefont{Vidal}} \bibnamefont{and} \bibinfo{author}{\bibfnamefont{R.~F.} \bibnamefont{Werner}}, \bibinfo{journal}{Physical Review A} \textbf{\bibinfo{volume}{65}}, \bibinfo{pages}{032314} (\bibinfo{year}{2002}).

\bibitem[{\citenamefont{Plenio}(2005)}]{Plenio:2005cwa}
\bibinfo{author}{\bibfnamefont{M.~B.} \bibnamefont{Plenio}}, \bibinfo{journal}{Phys. Rev. Lett.} \textbf{\bibinfo{volume}{95}}, \bibinfo{pages}{090503} (\bibinfo{year}{2005}), \eprint{quant-ph/0505071}.

\bibitem[{\citenamefont{Parez and Witczak-Krempa}(2024)}]{Parez:2023xpj}
\bibinfo{author}{\bibfnamefont{G.}~\bibnamefont{Parez}} \bibnamefont{and} \bibinfo{author}{\bibfnamefont{W.}~\bibnamefont{Witczak-Krempa}}, \bibinfo{journal}{Phys. Rev. Res.} \textbf{\bibinfo{volume}{6}}, \bibinfo{pages}{023125} (\bibinfo{year}{2024}), \eprint{2310.15273}.

\bibitem[{\citenamefont{Cardy}(2013)}]{cardy2013some}
\bibinfo{author}{\bibfnamefont{J.}~\bibnamefont{Cardy}}, \bibinfo{journal}{Journal of Physics A: Mathematical and Theoretical} \textbf{\bibinfo{volume}{46}}, \bibinfo{pages}{285402} (\bibinfo{year}{2013}).

\bibitem[{\citenamefont{Calabrese et~al.}(2011)\citenamefont{Calabrese, Cardy, and Tonni}}]{calabrese2011entanglement}
\bibinfo{author}{\bibfnamefont{P.}~\bibnamefont{Calabrese}}, \bibinfo{author}{\bibfnamefont{J.}~\bibnamefont{Cardy}}, \bibnamefont{and} \bibinfo{author}{\bibfnamefont{E.}~\bibnamefont{Tonni}}, \bibinfo{journal}{Journal of Statistical Mechanics: Theory and Experiment} \textbf{\bibinfo{volume}{2011}}, \bibinfo{pages}{P01021} (\bibinfo{year}{2011}).

\bibitem[{\citenamefont{Fisher}(1995)}]{fisher1995critical}
\bibinfo{author}{\bibfnamefont{D.~S.} \bibnamefont{Fisher}}, \bibinfo{journal}{Physical review b} \textbf{\bibinfo{volume}{51}}, \bibinfo{pages}{6411} (\bibinfo{year}{1995}).

\bibitem[{\citenamefont{Refael and Moore}(2004)}]{refael2004entanglement}
\bibinfo{author}{\bibfnamefont{G.}~\bibnamefont{Refael}} \bibnamefont{and} \bibinfo{author}{\bibfnamefont{J.~E.} \bibnamefont{Moore}}, \bibinfo{journal}{Physical review letters} \textbf{\bibinfo{volume}{93}}, \bibinfo{pages}{260602} (\bibinfo{year}{2004}).

\bibitem[{\citenamefont{{Christandl} and {Winter}}(2004)}]{2004JMP....45..829C}
\bibinfo{author}{\bibfnamefont{M.}~\bibnamefont{{Christandl}}} \bibnamefont{and} \bibinfo{author}{\bibfnamefont{A.}~\bibnamefont{{Winter}}}, \bibinfo{journal}{Journal of Mathematical Physics} \textbf{\bibinfo{volume}{45}}, \bibinfo{pages}{829} (\bibinfo{year}{2004}), \eprint{quant-ph/0308088}.

\bibitem[{\citenamefont{{Cerf} and {Adami}}(1997)}]{1997PhRvL..79.5194C}
\bibinfo{author}{\bibfnamefont{N.~J.} \bibnamefont{{Cerf}}} \bibnamefont{and} \bibinfo{author}{\bibfnamefont{C.}~\bibnamefont{{Adami}}}, \bibinfo{journal}{\prl} \textbf{\bibinfo{volume}{79}}, \bibinfo{pages}{5194} (\bibinfo{year}{1997}), \eprint{quant-ph/9512022}.

\bibitem[{\citenamefont{{Koashi} and {Winter}}(2004)}]{2004PhRvA..69b2309K}
\bibinfo{author}{\bibfnamefont{M.}~\bibnamefont{{Koashi}}} \bibnamefont{and} \bibinfo{author}{\bibfnamefont{A.}~\bibnamefont{{Winter}}}, \bibinfo{journal}{\pra} \textbf{\bibinfo{volume}{69}}, \bibinfo{eid}{022309} (\bibinfo{year}{2004}), \eprint{quant-ph/0310037}.

\bibitem[{\citenamefont{Greenberger et~al.}(1989)\citenamefont{Greenberger, Horne, and Zeilinger}}]{greenberger1989going}
\bibinfo{author}{\bibfnamefont{D.~M.} \bibnamefont{Greenberger}}, \bibinfo{author}{\bibfnamefont{M.~A.} \bibnamefont{Horne}}, \bibnamefont{and} \bibinfo{author}{\bibfnamefont{A.}~\bibnamefont{Zeilinger}}, in \emph{\bibinfo{booktitle}{Bell’s theorem, quantum theory and conceptions of the universe}} (\bibinfo{publisher}{Springer}, \bibinfo{year}{1989}), pp. \bibinfo{pages}{69--72}.

\bibitem[{\citenamefont{{Hayden} et~al.}(2011)\citenamefont{{Hayden}, {Headrick}, and {Maloney}}}]{2011arXiv1107.2940H}
\bibinfo{author}{\bibfnamefont{P.}~\bibnamefont{{Hayden}}}, \bibinfo{author}{\bibfnamefont{M.}~\bibnamefont{{Headrick}}}, \bibnamefont{and} \bibinfo{author}{\bibfnamefont{A.}~\bibnamefont{{Maloney}}}, \bibinfo{journal}{arXiv e-prints} \bibinfo{eid}{arXiv:1107.2940} (\bibinfo{year}{2011}), \eprint{1107.2940}.

\bibitem[{\citenamefont{Shi et~al.}(2020)\citenamefont{Shi, Dai, and Lu}}]{shi2020entanglement}
\bibinfo{author}{\bibfnamefont{B.}~\bibnamefont{Shi}}, \bibinfo{author}{\bibfnamefont{X.}~\bibnamefont{Dai}}, \bibnamefont{and} \bibinfo{author}{\bibfnamefont{Y.-M.} \bibnamefont{Lu}}, \bibinfo{journal}{arXiv preprint arXiv:2012.00040}  (\bibinfo{year}{2020}).

\bibitem[{\citenamefont{{Parez} and {Witczak-Krempa}}(2024)}]{2024arXiv240206677P}
\bibinfo{author}{\bibfnamefont{G.}~\bibnamefont{{Parez}}} \bibnamefont{and} \bibinfo{author}{\bibfnamefont{W.}~\bibnamefont{{Witczak-Krempa}}}, \bibinfo{journal}{arXiv e-prints} \bibinfo{eid}{arXiv:2402.06677} (\bibinfo{year}{2024}).

\bibitem[{\citenamefont{{Laflorencie}}(2005)}]{2005PhRvB..72n0408L}
\bibinfo{author}{\bibfnamefont{N.}~\bibnamefont{{Laflorencie}}}, \bibinfo{journal}{\prb} \textbf{\bibinfo{volume}{72}}, \bibinfo{eid}{140408} (\bibinfo{year}{2005}), \eprint{cond-mat/0504446}.

\bibitem[{\citenamefont{{Devetak} and {Winter}}(2005)}]{2005RSPSA.461..207D}
\bibinfo{author}{\bibfnamefont{I.}~\bibnamefont{{Devetak}}} \bibnamefont{and} \bibinfo{author}{\bibfnamefont{A.}~\bibnamefont{{Winter}}}, \bibinfo{journal}{Proceedings of the Royal Society of London Series A} \textbf{\bibinfo{volume}{461}}, \bibinfo{pages}{207} (\bibinfo{year}{2005}), \eprint{quant-ph/0306078}.

\bibitem[{\citenamefont{{Horodecki} et~al.}(2000)\citenamefont{{Horodecki}, {Horodecki}, and {Horodecki}}}]{2000PhRvL..85..433H}
\bibinfo{author}{\bibfnamefont{M.}~\bibnamefont{{Horodecki}}}, \bibinfo{author}{\bibfnamefont{P.}~\bibnamefont{{Horodecki}}}, \bibnamefont{and} \bibinfo{author}{\bibfnamefont{R.}~\bibnamefont{{Horodecki}}}, \bibinfo{journal}{\prl} \textbf{\bibinfo{volume}{85}}, \bibinfo{pages}{433} (\bibinfo{year}{2000}), \eprint{quant-ph/0003040}.

\bibitem[{\citenamefont{{Kudler-Flam} et~al.}(2025)\citenamefont{{Kudler-Flam}, {Narovlansky}, and {Sopenko}}}]{Long_Range_Entanglement_Figure4_2025}
\bibinfo{author}{\bibfnamefont{J.}~\bibnamefont{{Kudler-Flam}}}, \bibinfo{author}{\bibfnamefont{V.}~\bibnamefont{{Narovlansky}}}, \bibnamefont{and} \bibinfo{author}{\bibfnamefont{N.}~\bibnamefont{{Sopenko}}}, \emph{\bibinfo{title}{{Figure 4 data from the Long\_Range\_Entanglement repository}}}, \bibinfo{howpublished}{\url{https://github.com/jkudlerflam/Long_Range_Entanglement/blob/main/Figure4}} (\bibinfo{year}{2025}).

\bibitem[{\citenamefont{{Vitagliano} et~al.}(2010)\citenamefont{{Vitagliano}, {Riera}, and {Latorre}}}]{2010NJPh...12k3049V}
\bibinfo{author}{\bibfnamefont{G.}~\bibnamefont{{Vitagliano}}}, \bibinfo{author}{\bibfnamefont{A.}~\bibnamefont{{Riera}}}, \bibnamefont{and} \bibinfo{author}{\bibfnamefont{J.~I.} \bibnamefont{{Latorre}}}, \bibinfo{journal}{New Journal of Physics} \textbf{\bibinfo{volume}{12}}, \bibinfo{eid}{113049} (\bibinfo{year}{2010}), \eprint{1003.1292}.

\bibitem[{\citenamefont{{Ram{\'\i}rez} et~al.}(2014)\citenamefont{{Ram{\'\i}rez}, {Rodr{\'\i}guez-Laguna}, and {Sierra}}}]{2014JSMTE..10..004R}
\bibinfo{author}{\bibfnamefont{G.}~\bibnamefont{{Ram{\'\i}rez}}}, \bibinfo{author}{\bibfnamefont{J.}~\bibnamefont{{Rodr{\'\i}guez-Laguna}}}, \bibnamefont{and} \bibinfo{author}{\bibfnamefont{G.}~\bibnamefont{{Sierra}}}, \bibinfo{journal}{Journal of Statistical Mechanics: Theory and Experiment} \textbf{\bibinfo{volume}{2014}}, \bibinfo{eid}{10004} (\bibinfo{year}{2014}), \eprint{1407.3456}.

\bibitem[{\citenamefont{{Ram{\'\i}rez} et~al.}(2015)\citenamefont{{Ram{\'\i}rez}, {Rodr{\'\i}guez-Laguna}, and {Sierra}}}]{2015JSMTE..06..002R}
\bibinfo{author}{\bibfnamefont{G.}~\bibnamefont{{Ram{\'\i}rez}}}, \bibinfo{author}{\bibfnamefont{J.}~\bibnamefont{{Rodr{\'\i}guez-Laguna}}}, \bibnamefont{and} \bibinfo{author}{\bibfnamefont{G.}~\bibnamefont{{Sierra}}}, \bibinfo{journal}{Journal of Statistical Mechanics: Theory and Experiment} \textbf{\bibinfo{volume}{2015}}, \bibinfo{eid}{06002} (\bibinfo{year}{2015}), \eprint{1503.02695}.

\bibitem[{\citenamefont{{Rodr{\'\i}guez-Laguna} et~al.}(2017)\citenamefont{{Rodr{\'\i}guez-Laguna}, {Dubail}, {Ram{\'\i}rez}, {Calabrese}, and {Sierra}}}]{2017JPhA...50p4001R}
\bibinfo{author}{\bibfnamefont{J.}~\bibnamefont{{Rodr{\'\i}guez-Laguna}}}, \bibinfo{author}{\bibfnamefont{J.}~\bibnamefont{{Dubail}}}, \bibinfo{author}{\bibfnamefont{G.}~\bibnamefont{{Ram{\'\i}rez}}}, \bibinfo{author}{\bibfnamefont{P.}~\bibnamefont{{Calabrese}}}, \bibnamefont{and} \bibinfo{author}{\bibfnamefont{G.}~\bibnamefont{{Sierra}}}, \bibinfo{journal}{Journal of Physics A Mathematical General} \textbf{\bibinfo{volume}{50}}, \bibinfo{eid}{164001} (\bibinfo{year}{2017}), \eprint{1611.08559}.

\bibitem[{\citenamefont{Movassagh and Shor}(2016)}]{movassagh2016supercritical}
\bibinfo{author}{\bibfnamefont{R.}~\bibnamefont{Movassagh}} \bibnamefont{and} \bibinfo{author}{\bibfnamefont{P.~W.} \bibnamefont{Shor}}, \bibinfo{journal}{Proceedings of the National Academy of Sciences} \textbf{\bibinfo{volume}{113}}, \bibinfo{pages}{13278} (\bibinfo{year}{2016}).

\bibitem[{\citenamefont{Bravyi et~al.}(2012)\citenamefont{Bravyi, Caha, Movassagh, Nagaj, and Shor}}]{bravyi2012criticality}
\bibinfo{author}{\bibfnamefont{S.}~\bibnamefont{Bravyi}}, \bibinfo{author}{\bibfnamefont{L.}~\bibnamefont{Caha}}, \bibinfo{author}{\bibfnamefont{R.}~\bibnamefont{Movassagh}}, \bibinfo{author}{\bibfnamefont{D.}~\bibnamefont{Nagaj}}, \bibnamefont{and} \bibinfo{author}{\bibfnamefont{P.~W.} \bibnamefont{Shor}}, \bibinfo{journal}{Physical review letters} \textbf{\bibinfo{volume}{109}}, \bibinfo{pages}{207202} (\bibinfo{year}{2012}).

\bibitem[{\citenamefont{Zhang et~al.}(2016)\citenamefont{Zhang, Ahmadain, and Klich}}]{zhang2016quantum}
\bibinfo{author}{\bibfnamefont{Z.}~\bibnamefont{Zhang}}, \bibinfo{author}{\bibfnamefont{A.}~\bibnamefont{Ahmadain}}, \bibnamefont{and} \bibinfo{author}{\bibfnamefont{I.}~\bibnamefont{Klich}}, \bibinfo{journal}{arXiv preprint arXiv:1606.07795}  (\bibinfo{year}{2016}).

\end{thebibliography}

\end{document}